\documentclass[twocolumn]{aastex631}

\shorttitle{}
\shortauthors{Cohen et al.}

\graphicspath{{./}{figures/}}

\begin{document}

\title{Three-dimensional, Time-dependent MHD Simulation of Disk-Magnetosphere-Stellar Wind Interaction in a T Tauri, Protoplanetary System}

\correspondingauthor{Ofer Cohen}
\email{ofer\_cohen@uml.edu}

\author[0000-0003-3721-0215]{Ofer Cohen}
\affiliation{Lowell Center for Space Science and Technology, University of Massachusetts Lowell, 600 Suffolk Street, Lowell, MA 01854, USA}

\author[0000-0002-8791-6286]{Cecilia Garraffo}
\affiliation{Harvard-Smithsonian Center for Astrophysics, 60 Garden Street, Cambridge, MA 02138, USA}

\author[0000-0002-0210-2276]{Jeremy J. Drake}
\affiliation{Harvard-Smithsonian Center for Astrophysics, 60 Garden Street, Cambridge, MA 02138, USA}

\author[0000-0002-5688-6790]{Kristina Monsch}
\affiliation{Harvard-Smithsonian Center for Astrophysics, 60 Garden Street, Cambridge, MA 02138, USA}

\author[0000-0002-6118-0469]{Igor V. Sokolov}
\affiliation{Department of Climate and Space Sciences and Engineering, University of Michigan, 2455 Hayward, Ann Arbor, MI 48109, USA}

\author[0000-0001-5052-3473]{Juli\'{a}n D. Alvarado-G\'{o}mez}
\affiliation{Leibniz Institute for Astrophysics Potsdam, An der Sternwarte 16, 14482 Potsdam, Germany}

\author[0000-0002-5456-4771]{Federico Fraschetti}
\affiliation{Dept. of Planetary Sciences-Lunar and Planetary Laboratory, University of Arizona, Tucson, AZ, 85721, USA}
\affiliation{Harvard-Smithsonian Center for Astrophysics, 60 Garden Street, Cambridge, MA 02138, USA}

%%%%%% Abstract %%%%%%%%%%

\begin{abstract}

We present a three-dimensional, time-dependent, MHD simulation of the short-term interaction between a protoplanetary disk and the stellar corona in a T Tauri system. The simulation includes the stellar magnetic field, self-consistent coronal heating and stellar wind acceleration, and a disk rotating at sub-Keplerian velocity to induce accretion. We find that initially, as the system relaxes from the assumed initial conditions, the inner part of the disk winds around and moves inward and close to the star as expected. However, the self-consistent coronal heating and stellar wind acceleration build up the original state after some time, significantly pushing the disk out beyond $10R_\star$. After this initial relaxation period, we do not find clear evidence of a strong, steady accretion flow funneled along coronal field lines, but only weak, sporadic accretion. We produce synthetic coronal X-ray line emission light curves which show flare-like increases that are not correlated with accretion events nor with heating events. These variations in the line emission flux are the result of compression and expansion due to disk-corona pressure variations. Vertical disk evaporation evolves above and below the disk. However, the disk - stellar wind boundary stays quite stable, and any disk material that reaches the stellar wind region is advected out by the stellar wind.

\end{abstract}


%%%%%% Introduction %%%%%%%%%%

\section{Introduction} \label{sec:intro}

The interaction between the inner parts of protoplanetary disks and the stellar magnetosphere plays a major role in the overall angular momentum transfer in the system, which controls the spin up of the star, and the long-term disk stability. The theory of protoplanetary disks traditionally accounts for an accretion of mass in the inner parts of the disk, which is driven by turbulence/viscosity instabilities. The mass accretion onto the star takes angular momentum from the disk. Thus, the loss of angular momentum requires a gain of angular momentum in other parts of the disk in order to conserve angular momentum and to keep the disk stable over an estimated lifetime of about 10 million years \citep[see review in][]{2020apfs.book.....A}. The magnitude or level of the turbulence in the disk is characterized by a free parameter, $\alpha$, \citep[the ``$\alpha$-disk model",][]{1973A&A....24..337S}, which can have a magnitude over a number of orders of magnitudes ($10^{-5}-10^{-1}$) and it is still poorly constrained by observations. The physical mechanism for the turbulence can be obtained from a Hydrodynamic prescription, or Magnetohydrodynamic (MHD) prescription \citep[Magneto-rotational Instability or MRI,][]{1991ApJ...376..214B}. 

In the quasi-static, long-term scenario, the innermost part of the protoplanetary disk is expected to be truncated by the stellar magnetosphere, so the disk material accretes along the magnetospheric magnetic field lines onto the star \citep[e.g.,][]{2012AN....333....4H,2023arXiv230111628T}. This picture has allowed estimation of mass and angular momentum transfer rates through the process, and the long-term amount of disk accreting mass and stellar spin up \citep{2020apfs.book.....A}. 

The ideal picture of a steady accretion through the large magnetospheric loops has been revised in more recent studies, which have looked into the opening of the stellar magnetic field lines by the rotating disk. These studies show a significant modification of the angular momentum transfer between the disk and the stellar corona compared to the picture of steady accretion along the assumed dipolar stellar field lines \citep{2000MNRAS.317..273A,2008ApJ...678.1109M,2009A&A...508.1117Z,2017ApJ...845...75B,2017A&A...600A..75B,2020A&A...643A.129P,2022ApJ...929...65I}.

The discussion above refers to the long-term, quasi-static disk-corona interaction. The short-term interaction in these regions may be much more dynamic, and the picture of steady accretion may not hold. Specifically, it is not clear how the significantly ionized disk material can penetrate the closed magnetic field lines of the stellar magnetosphere without a steady magnetic reconnection in the disk-magnetosphere boundary. Since magnetic reconnection is a small-scale effect, such a steady reconnection over timescales of a full disk rotation (hours to days) can only be maintained in an idealized, unrealistic scenario. 

The short-term dynamic interaction between the disk and the stellar magnetosphere can drive a set of processes, including magnetic reconnection \citep[see recent review by][]{2022LRSP...19....1P}, magnetic field line inflation \citep[e.g.,][]{1990RvMA....3..234C}, and propeller-driven outflows, on top of the disk wind and photoevaporation from the disk at greater distances from the star \citep[see review by][]{2015SSRv..191..339R}. These dynamic, short-term processes bring the picture of steady disk accretion through magnetospheric field lines into question. Instead, it seems more likely that disk accretion happens in an episodic manner \citep[see review by][]{2014prpl.conf..387A}. This behavior has been demonstrated in a number of dynamic, multi-dimensional simulations  \citep[summarized in][]{2015SSRv..191..339R} and also \citep[e.g.,][]{2013A&A...559A.127O,2019A&A...624A..50C,2020A&A...644A..74V,2020MNRAS.495.3494Z,2022A&A...667A..17M,2022ApJ...941...73T}. It is therefore desirable to relate the short-term disk-magnetosphere dynamics, which can be probed by observations, to the parameters that describe the long-term disk evolution, i.e., the average mass and angular momentum loss rates.

Observations of disk-magnetosphere interaction in Classical T Tauri Stars (CTTS) have shown temporal variability of the order of days, with a non-stationary accretion from the disk onto the magnetosphere and different sources of the variability \citep[e.g.,][]{2003A&A...409..169B,2010A&A...519A..88A,2012A&A...541A.116A,2014AJ....147...82C,2018ApJ...852...56G,2023arXiv230111693D}. Observational constraints for the mass accretion rates in CTTS are highly uncertain, generally spanning many orders of magnitudes in the range of $10^{-11}-10^{-7}~M_\odot~yr^{-1}$ \citep[see summary by][]{2017ASPC..511...28B}. Most of the above studies conclude that the mass accretion rate falls somewhere in the middle of that range, around $10^{-9}-10^{-8}~M_\odot~yr^{-1}$. It is important to note that these estimates also rely on theoretical relations between the accreting mass and the observed luminosities of different lines, which depend, e.g., on an assumed inner truncation radius of the disk. For example, \cite{2012A&A...548A..56R} and \cite{2014A&A...561A...2A} assumed an inner radius of $5R_\star$ or less based on \cite{2007ApJ...664..975J}. We note that it is possible that the disk inner radius is greater than that, since these estimates did not account for the stellar wind, but only the stellar magnetic field strength (which by itself is uncertain). It is also worth mentioning that the scatter of the observations around the derived scaling laws is typically quite large.

%Even the more recent models for disk-magnetosphere interaction did not fully account for the properties of the stellar corona and stellar wind (see a detailed discussion in Section~\ref{sec:discussion}). 
Important aspects of the star-disk interaction problem are the properties of the stellar corona and stellar wind. It is now known from our own Sun that magnetic energy plays a significant role in heating the corona and accelerating the wind. The particular way this energy is dissipated into the plasma is not yet fully understood, but it can be via either Alfv\'en wave dissipation, particle resonance, or small-scale magnetic reconnection \citep[or all three, see e.g.,][]{2005psci.book.....A,2018LRSP...15....4G}. The ``body" with which the disk interacts is therefore not static, but is an energized body. The coronal plasma is hot (several MK), the stellar wind is accelerated to a few hundred $km~s^{-1}$, and the stellar magnetic field is not potential as it is stretched and opened up by the wind. To date, no numerical model for the disk-magnetosphere interaction has accounted for the effect of a self-consistent hot corona and accelerated stellar wind. The magnetospheric plasma is typically specified with some initial and boundary conditions, where a weak thermally driven wind \citep{1958ApJ...128..664P} may develop due to the buildup of a pressure gradient between the inner boundary and the simulation domain \citep[e.g.,][]{2008ApJ...678.1109M,2022ApJ...941...73T}. It is therefore important to account for the impact of the energized corona on the ``disk-corona interaction", not ``disk-magnetosphere interaction", where one should expect that the interaction will release some of the stored energy. 

Another aspect has been missing from disk-magnetosphere models. The stellar wind in the regions above and below the disk is super-Alfv\'enic. Thus, it is important to investigate how the sub-Alfv\'enic disk region interacts with this surrounding super-Alfv\'enic stellar wind, and whether it can impact the development of disk winds and disk photoevaporation. A main point to make in this context is that in models which do not have additional energy and momentum terms in their equations, the initially energized Parker wind may be suppressed by the interaction with the disk as well as by the stellar potential field. This is due to the fact that the steady state of these models represents a minimum energy, relaxed state. 
%In the case of our model, the wind and corona are continuously energized through the additional terms, so the interaction with the disk is expected to be different. 

In this paper we explore the disk-corona interaction using a three-dimensional, time-dependent MHD model that includes the stellar corona, stellar wind, and a marginally-stable Keplerian disk. The wind and corona are continuously energized through the additional terms, and the interaction with the disk is expected to be different to that seen in eaerlier studies. We describe the model setup in Section~\ref{sec:model}, and present our results in Section~\ref{sec:results}. We then discuss our finding in Section~\ref{sec:discussion} and draw our final conclusions in Section~\ref{sec:summary}.

%%%%%% Model Description %%%%%%%%%%

\section{Model Description} \label{sec:model}

\subsection{Background Stellar Wind Model} \label{sec:SWmodel}

Here, we use the Alfv\'en Wave Solar Model \citep[AWSoM][]{2014ApJ...782...81V} to obtain the background stellar corona and stellar wind solution. AWSoM solves the three-dimensional non-ideal MHD equations, including additional terms in the momentum and energy equations that account for the wind acceleration and coronal heating, respectively. In addition, AWSoM accounts for coronal thermodynamics, including electron heat conduction and radiative cooling. AWSoM can capture the stellar chromosphere and transition region. However, this requires a very small grid size near the inner boundary, which makes the time step very small. 

To overcome the issue of small time step, the AWSoM-R \citep{2021ApJ...908..172S} version was introduced. AWSoM-R assumes that the magnetic field in between the chromosphere and the low corona is potential and the magnetic field lines are vertical to the surface. The model uses a thread of one dimensional thermodynamic models along selected field lines to cover the three-dimensional domain in the chromosphere and transition region (1-1.05$R_\star$), replacing the the three-dimensional solution by AWSOM. This allows the AWSoM domain to start at a slightly higher altitude above the transition region (the low corona), where the grid size can be larger, and so does the timestep. 

AWSoM is driven by the boundary condition of the stellar magnetic field (magnetic field maps or ``magnetograms"), which can be obtained observationally or analytically. AWSoM has many advanced and versatile options regarding the numerical schemes, controlling $\nabla\cdot\mathbf{B}$, adaptive mesh refinement, time stepping, and implicit/explicit options. We refer the reader to \cite{1999JCoPh.154..284P,2012JCoPh.231..870T,2014ApJ...782...81V} for the model's complete description and capabilities.

AWSoM has been used to simulate the coronae and winds of many solar analogs in the context of the space environment of exoplanets \citep[e.g.,][]{2011ApJ...733...67C,2016A&A...588A..28A,2016ApJ...833L...4G,2017ApJ...843L..33G,2018PNAS..115..260D,2019ApJ...884L..13A} and stellar evolution \citep[e.g.,][]{2014ApJ...783...55C,2016A&A...595A.110G,2018ApJ...862...90G}. Here, we choose a Sun-like star, with solar mass, radius and rotation period of 90 days (essentially no rotation). The simulation is performed in the frame of reference that is rotating with the star, so the dynamics is driven by the Keplerian rotation of the disk with respect to the stellar corona and stellar wind. While classical T Tauri stars may have a rotation period of 10 days or less, we expect that any dynamics captured by our simulation would be enhanced with an enhanced stellar rotation. In this initial study, we use a $100~G$ dipole stellar field tilted by 45$^\circ$ to enhance disk-magnetosphere interaction. A higher-order, complex stellar (or solar) field can be easily implemented in future studies, even though the complexity would probably make it harder to interpret the results. 

\begin{figure*}[h!]
\centering
\includegraphics[width=5.in]{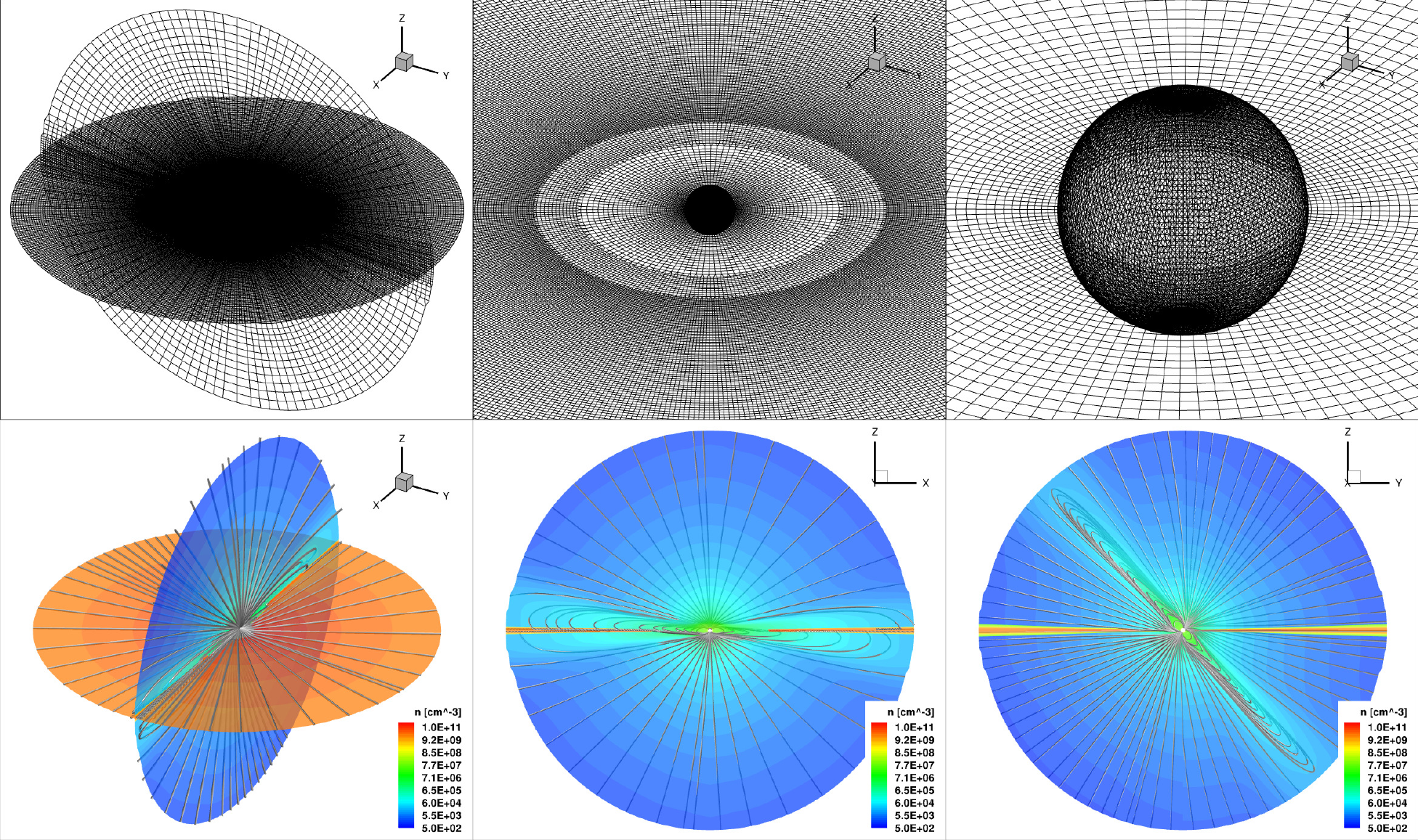}
\caption{Top: the grid structure in the simulation displayed on a sphere of $r=1.1R_\star$, and on the $z=0$, and $x=0$ (top-left only) slices. The full domain is shown on the top-left panel, while the top-middle, and top-right panels show a zoom-in view at different levels, respectively. Bottom: the initial condition of the simulation. Bottom-left panel shows a three-dimensional view of the $z=0$ and $y=0$ slices colored with number density contours, and selected magnetic field lines are shown in grey. Bottom-middle and bottom-right panels show a similar display of the $y=0$ and $x=0$ slices, respectively.}
\label{fig:Grid}
\end{figure*}

We use a spherical grid stretched in the radius with a smallest grid cell size of $0.02R_\star$ (angular resolution of 0.07 degree in longitude and latitude) near the inner boundary. We also refine the grid around the equatorial plane to capture the thin disk, with about 6-10 grid cells covering the disk in different regions. The grid structure is shown in the top panels of Figure~\ref{fig:Grid}. Overall, the simulations presented here are very expensive and time consuming, with a total of 5.6 million grid cells in the simulation. 

In order to see whether this grid definition is sufficient, we performed a simulation with double the number of grid cells across the disk. We found that the increased resolution did not change the results much after 50 hours (the overall important trends were the same). The time step, however, was much smaller and, with the total number of cells being significantly larger, simulations with the higher resolution would have been only marginally feasible. 

\subsection{Disk Model} \label{sec:Diskmodel}

\subsubsection{Initial Hydrostatic Disk}\label{HydrostaticDisk}

While AWSoM uses a Spherical grid, it uses Cartesian terminology to describe a particular location on the grid. We adopt a spherical description for the disk with the standard coordinate transformation of:
\begin{eqnarray}
&r=\sqrt{x^2+y^2+z^2}& \\
&\theta=\cos^{-1}{\left( \frac{z}{r} \right)}& \nonumber\\
&\phi=\tan^{-1}{\left( \frac{y}{x} \right)}& \nonumber
\end{eqnarray}
for radius, latitude, and longitude, respectively. We also define the equatorial radius, $R_e=\sqrt{x^2+y^2}$. 

For our initial disk, we choose a maximum disk temperature of $T_0=500~K$ and we specify the decrease of the disk temperature with the radius as $T(R_e)=T_0/R_e$ \citep{2020apfs.book.....A}. The Keplerian angular velocity is defined as $\Omega_k(r)=\sqrt{2GM_\star/r^3}$, with $G$ being the gravitational constant, and $M_\star$ is the stellar mass. Similarly, the sound speed is defined as $C_s=\sqrt{kT/m_p}$, with $k$ being the Boltzmann constant, and $m_p$ being the proton mass. Using these definitions, we can calculate the disk scale height, $h=C_s/\Omega_k$, which increases with radius. Once the disk's scale height is defined, it is used to define the hydrostatic profiles for the density and pressure for the disk's initial condition as follows:
\begin{eqnarray}
&\rho(R_e,z)=\frac{1}{R_e}\rho_0 \exp{\left( -|z|/h \right)}& \\
&p(R_e,z)= \rho(R_e,z)C^2_s& \nonumber
\end{eqnarray}
In the radial direction, the disk equatorial density drops from its maximum value as $1/R_e$, and it drops exponentially with in the vertical (z) direction. The maximum density, $\rho_0$, is chosen to be $10^{13}~[cm^{-3}]$, based on values used in previous simulations \citep[e.g., ][]{2002ApJ...576L..53K}. We choose an upper limit value so that the disk is heavy enough to have some initial accretion.

The initial velocity in the disk is set to Keplerian, with the exception of its inner parts (see Section~\ref{DiskInstability} below). In many previous simulations, the magnetic field inside the disk is prescribed. However, our goal here is to capture the interaction of the disk with the surrounding magnetic field of the stellar wind (see discussion in Section~\ref{NonVerticalB}). Therefore, we initially set the magnetic field inside the disk to zero, and we let the disk's magnetization evolve with time as it interacts with the surrounding stellar and wind magnetic fields (this happens quickly). The additional terms associated with the Alfv\'en wave heating are also set to zero in the initial disk. With that, all the necessary MHD parameters are defined for the disk's initial condition.

Once the initial disk structure is defined, we overwrite the coronal and stellar wind steady-state solution with the disk solution in portions of the domain. In order to avoid an arbitrary choice of the disk-magnetosphere truncation radius, we overwrite the the steady-state wind solution if the thermal pressure of the initial hydrostatic disk in some location is greater than the thermal pressure in the steady-state stellar wind at that location. This conditions provides a more consistent initial location of the disk, and it also provides a more gradual transition between the background stellar wind solution and the new solutions with the imposed disk. Using the total pressure instead of the thermal pressure as our overwriting condition made almost no difference, since the thermal pressure dominates the initial disk. The initial density and magnetic field are shown in the bottom panels of Figure~\ref{fig:Grid}.

\subsubsection{Sub-Keplerian Motion}\label{DiskInstability}

Disk instability requires the introduction of some diffusion/viscosity processes that reduce the Keplerian velocity via turbulence or MRI. In the case of our global model and the resolution used, these small-scale processes cannot be captured in a self-consistent manner. For example, MRI requires about 20 grid cells across the disk in order to be captured \citep{2013ApJ...772..102H}. Our simulations only marginally have such resolution so we do not expect to rely on a self-consistent viscosity. Instead, we modify the Keplerian velocity by a factor, $\eta$, which reduces the Keplerian velocity at the inner parts of the disk. This factor is defined as   
\begin{equation}
\eta(R_e)=1-0.75*\exp{(-R^2_e/10^2)},
\end{equation}
with $u_\phi(R_e)=\eta(R_e)u_k$, and $u_k(r)=\sqrt{2GM_\star/r}$. Thus, $\eta$ causes the Keplerian velocity to decrease going inwards toward the star, resulting in the most significant decrease inside $R_e=10R_\star$. This renders the disk marginally unstable and provokes accretion. 

\subsection{Scope and Limitations of the Numerical Study} \label{sec:Shortterm}

The goal of this global MHD study is to investigate the {\it large-scale} interaction between the stellar corona and the background stellar wind with the inner parts of the disk over a timescale of about 100 hours. By definition, our simulation cannot resolve small-scale processes of disk MHD dynamics which are studied in disk local box simulations, such as magnetic field amplification and the rise of self-consistent MRI and turbulence. The main goals of our study are to investigate: 1) how does the inclusion of self-consistent heating and acceleration terms in the MHD equations affect disk accretion and angular momentum transfer between the disk and the stellar corona; 2) what is the nature of the boundary between the disk and the super-Alfv\'enic stellar wind; and 3) how does the disk affects the global structure of the stellar field and the interplanetary field. 

We run the simulation for 108 hours, which is slightly longer than one Keplerian rotation at $r=10R_\star$ (the vicinity of the inner disk edge). This time period is sufficient to capture the short-term disk-corona interaction features, such as magnetic winding by the disk, and accretion episodes due to magnetic disconnection - the large-scale change in magnetic topology, distinguished from the small-scale magnetic ``reconnection". We also capture the mass and angular momentum loss rate within our time frame (See Section~\ref{MdotJdot}). 

The long-term interaction, which captures potential periodicity in the accretion episodes, and the long-term, evolutionary changes in mass and angular momentum loss rates, requires the simulation of multiple Keplerian rotations. Since our initial simulation is computationally expensive, we choose to focus here on the short term dynamics, and we leave the longer-term investigation for future studies. 

%%%%%% Results %%%%%%%%%%

\section{Results} \label{sec:results}

The initial behaviour of the disk-corona system is characterized by a relaxation from the initial starting conditions. These initial conditions by necessary expedient do not represent a physically realistic state of the system; i.e., they would not occur in a real star-disk situation. However, we consider the initial behaviour of the simulation still physically interesting and include this phase in the presentation of the results below. 

We consider the relaxation phase, during which magnetic field lines are wound by differential rotation, and the disk gravitational infall begins to balance against the coronal magnetic field and wind dynamic pressures, to end approximately 25 hours into the simulation.

\subsection{Global Structure of the Disk-corona Interaction Region}\label{GlobalStructure}

\begin{figure*}[h!]
\centering
\includegraphics[width=5in]{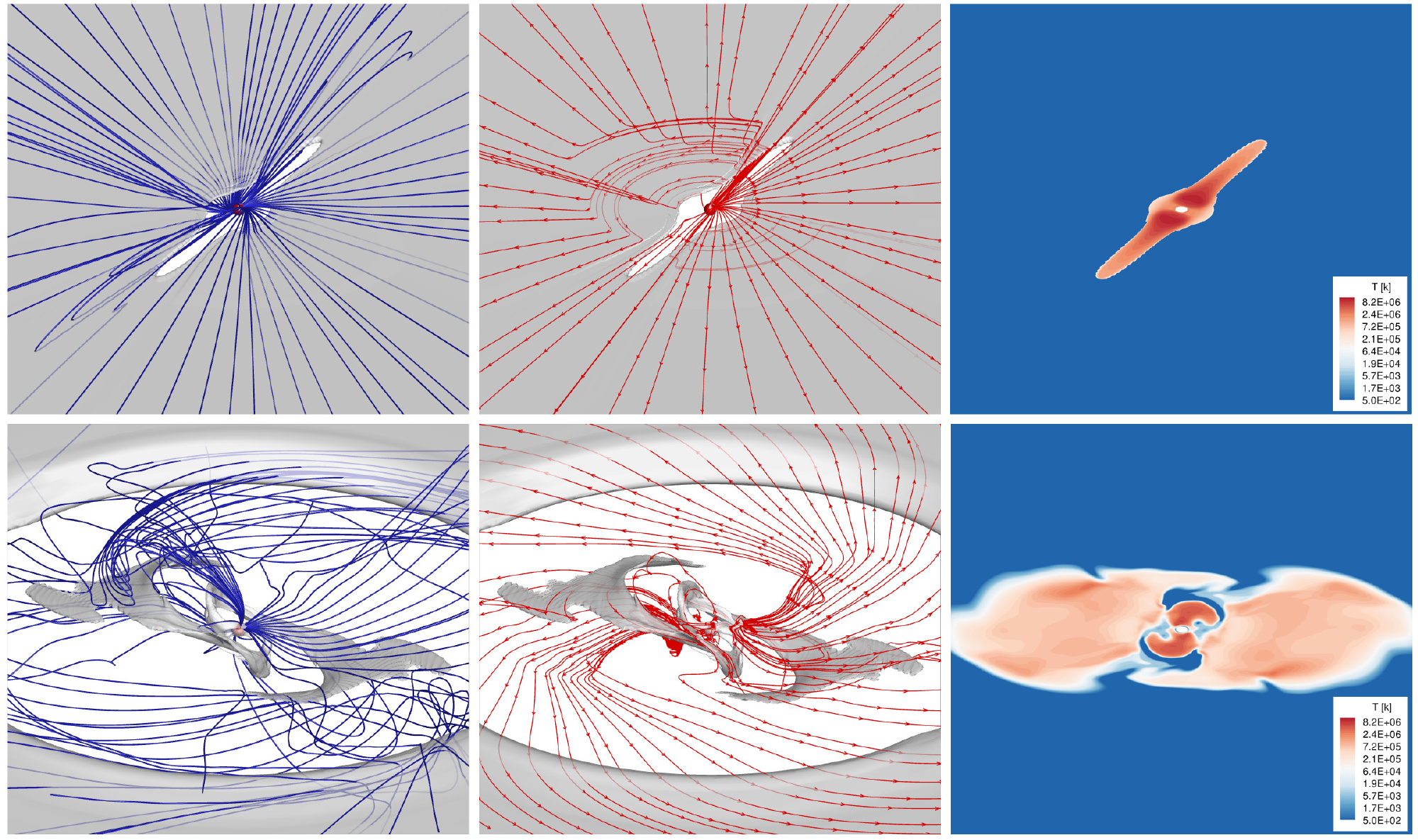}
\caption{The three-dimensional structure of the disk-corona inner region as seen from 30 degrees above the equator (similar to Figure~\ref{fig:Grid} left panels). The left panel shows a white iso-surface of $n=10^{10}~cm^{-3}$, which represents the higher density region of the disk. The two initial gaps in the iso-surface are due to the tilt of the stellar dipole field. The blue lines mark selected magnetic field lines that cover the displayed region. The middle panel shows the same iso-surface with red-colored velocity streamlines. In the right panel, we show the equatorial plane colored with contours of the temperature, where blue regions mark the cold disk material and red regions mark the hot corona. These are sample snapshots for $t=0$ (top panels) and $t=108h$ (bottom panels). Similar figures for all time steps are available as an animation.}
\label{fig:global_3d}
\end{figure*}

Figure~\ref{fig:global_3d} illustrates the global structure of the simulation at t=0h. The complete evolution of the simulation for this display is provided as an animation. It shows the inner part of the disk on an iso-surface for a number density of $10^{10}~cm^{-3}$, along with selected magnetic field line rods (left panel), and velocity streamlines (middle panel). The righ panel shows the equatorial plane colored with temperature contours. The field lines and streamlines are selected so that they cover large part of the displayed region. It can be seen that the initial disk-corona boundary, which is defined by the initial thermal pressure of the disk, is not circular, but it has two ``wing" gaps due to the non-symmetric tilted stellar dipole field. 

The global structure view shows a number of features. First, the coronal magnetic field, which is initially zero in the disk is being warped around by the rotational motion of the disk. After one Keplerian rotation, the field in the disk is almost purely azimuthal, and the {\it radial} stellar wind field is affected by this warping (see Section~\ref{NonVerticalB} for a further discussion on the consequences of this result). The second notable feature in the time-evolution of the disk-corona interaction is the expansion of the corona from the initial state, pushing the disk boundary outward. Even though it is set by a pressure condition, the initial disk-corona boundary is not expected to be balanced, as noted earlier. Due to the evolution of the additional momentum and energy terms in the model, parts of the stellar corona, which are being overwritten by the initial disk, redevelop, pushing the inner region of the initial disk away from the star. This expansion of the disk-corona boundary converges to a more steady location, which is quite far from the star, at around $15-20R_\star$.

The density iso-surfaces in the left and middle panels of Figure~\ref{fig:global_3d} are colored with the sign of the radial velocity. Grey represents outwards motion of the plasma, while black represents inward (accreting) motion. Our simulation shows an initial warping of the inner parts of the disk towards the star, leading to parts of the disk to get closer to the stellar surface. However, after some time, these closer parts of the disk are being pushed away. We discuss this pattern and the overall disk accretion in Section~\ref{AccretionPattern}.

\subsection{Radial Pressure Balance between the Disk and the Stellar Corona}\label{RadialPressureBalance}

\begin{figure*}[h!]
\centering
\includegraphics[width=6.in]{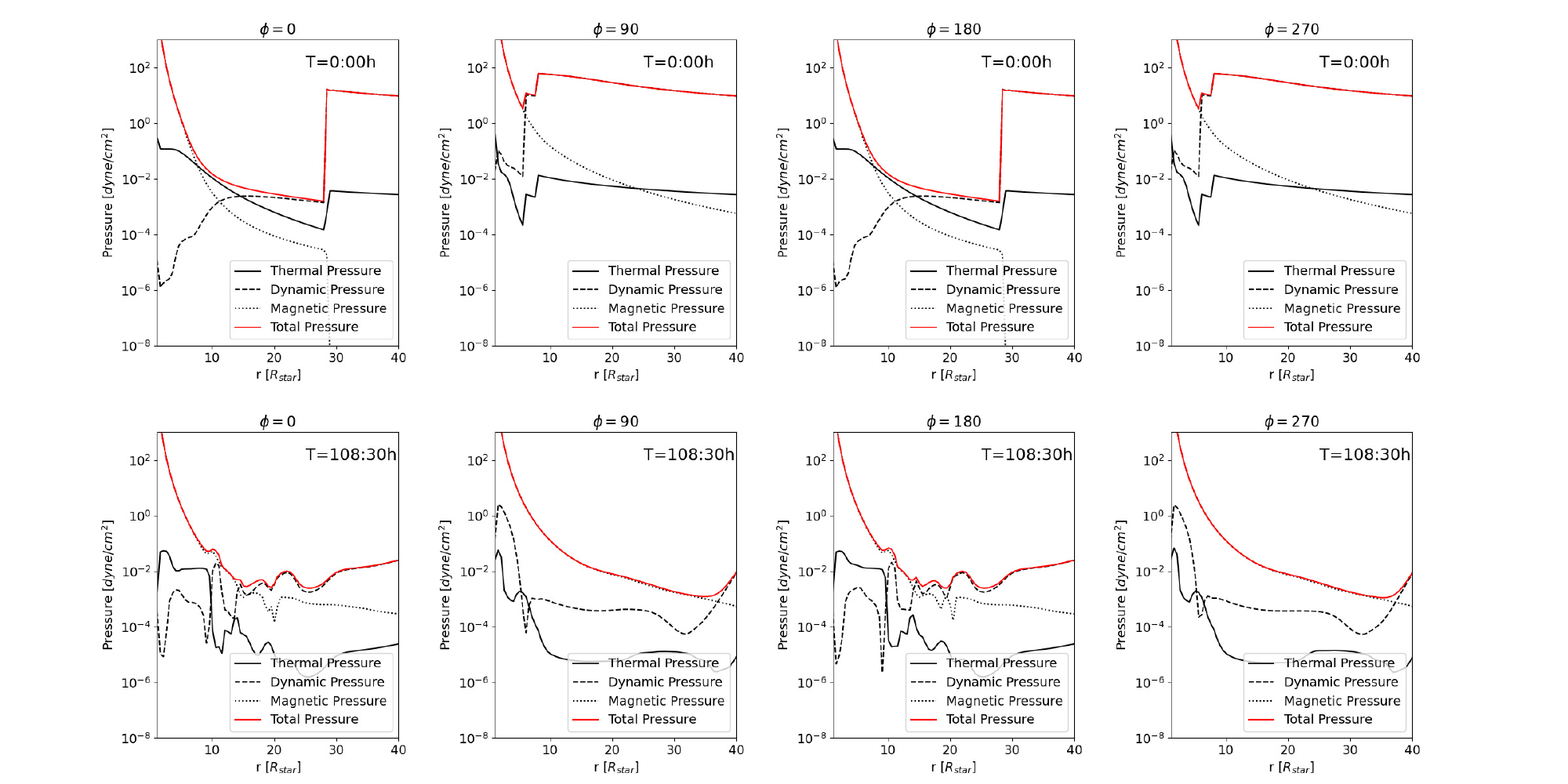}
\caption{Temporal evolution of the radial distribution of the different pressure components as a function of distance from the star. The pressure components are the thermal, dynamic, magnetic and total pressure. Plots show the radial distribution along extracted lines in the equatorial plane at four longitude: 0 (left column), 90 (second column), 180 (third column), and 270 (right column) degrees. These are sample snapshots for $t=0$ (top panels) and $t=108h$ (bottom panels). Similar figures for all time steps are available as an animation.}
\label{fig:P_4Q}
\end{figure*}

Figure~\ref{fig:P_4Q} shows the radial variation of the thermal, dynamic, magnetic, and total pressure in the equatorial plane. The figure shows the different pressure components at t=0h as a function of radius and for four different longitudes: 0, 90, 180, and 270 degrees. The complete evolution of the simulation for this display is provided as an animation. It can be seen that initially during the relaxation phase the total pressure is dominated by the disk's thermal pressure and by the disk's Keplerian dynamic pressure very close to the star. The magnetic pressure in the disk is very low due to the initial setting of a zero magnetic field inside the disk. As the simulation evolves, magnetic pressure builds up inside the disk due to the azimuthal winding of the field. Additionally, the coronal magnetic, thermal, and dynamic pressures, which were overwritten by the initial disk start to recover, pushing the disk away from the star. The main result from Figure~\ref{fig:P_4Q} is that in the inner regions, where the stellar corona was replaced by the initial disk, the corona recovers, pushing the disk away beyond $10~R_\star$.

\subsection{Vertical Evolution of the Disk}\label{VerticalEvolution}

\begin{figure*}[h!]
\centering
\includegraphics[width=5.in]{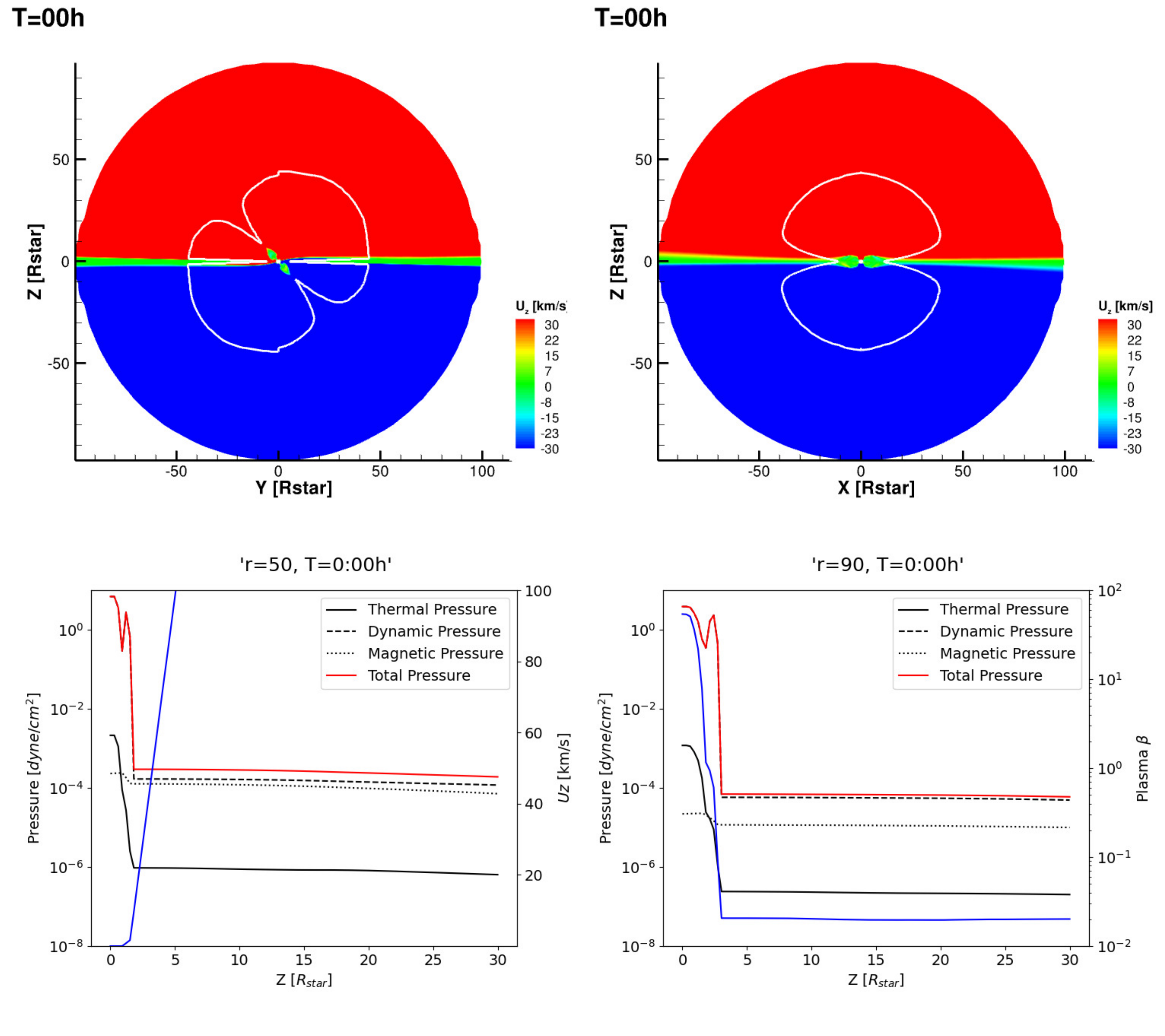}
\caption{Left and second columns: $x=0$ and $y=0$ meridional cuts colored with the vertical ($u_x$) component of the velocity. The solid white line represents location where the Alfv\'enic Mach number equals one. Third and right columns: vertical structure of the simulation pressure as a function of height above the equatorial plane. The vertical lines are extracted at $r=50R_\star$ (third panel) and $r=90R_\star$ (right panel). The pressure components are the thermal, dynamic, magnetic and total pressure. Also shown are the value of the vertical component of the velocity (third panel) and the plasma $\beta$ (right panel). This is a sample snapshot at $t=0$, similar figures for all time steps are available as an animation.}
\label{fig:VerticalP}
\end{figure*}

Figure~\ref{fig:VerticalP} shows the evolution of the disk's vertical structure at t=0h. The complete evolution of the simulation for this display is provided as an animation. It shows a vertical one dimensional extraction from the disk center ($z=0$) in to the stellar wind (up to $z=30R_\star$) in the Y-Z plane. The two extractions are at $r=50R_\star$ (middle of the simulated disk), and at $r=90R_\star$ (close to the outer edge of the simulation domain). The line plots show the different components of the pressure - thermal, dynamic, magnetic, and total. The plot for $r=50R_\star$ also shows the vertical ($v_z$) velocity, while the plot for $r=90R_\star$ also shows the plasma $\beta$. Figure~\ref{fig:VerticalP} also shows X-Z and Y-Z meridional slices colored with the vertical velocity. Also shown is the location of the Alfv\'en surface in solid white line.

The line plots in Figure~\ref{fig:VerticalP} show a clear separation between the disk and the stellar wind, represented by plasma $\beta>1$ inside the disk and plasma $\beta<1$ in outside of the disk. It is also represented by very low vertical velocity inside the disk, and a dramatic change to a very high (a few hundred $km~s^{-1}$) $v_z$ in the stellar wind. This is because the overall radial stellar wind having a vertical component due to the tilt of the stellar dipole. Initially, the disk expands vertically but this expansion slows down at some point. The line plots show that the vertical disk-wind (disk evaporation) evolves at the boundary region between the disk and the stellar wind, with velocities around $5-10~km~s^{-1}$. The meridional evolution does not show any indication that the evaporating disk material strongly penetrates into the stellar wind region. Instead, it appears that it is advected outwards by the stellar wind. %when it gets there. 
The total pressure is dominated by the dynamic (Keplerian) pressure in the disk, and by the dynamic and magnetic pressure in the stellar wind (the two are comparable). The thermal pressure seems to play a minor role inside the disk, and no role in the stellar wind.

The meridional cut plots show that even when the disk-wind boundary reaches an equilibrium state, the boundary is not static, and we start to see Kelvin-Helmholtz (KH) features, which are the result of the two interacting flows. These features are advected out by the stellar wind after they form. Despite the dynamic interaction at the disk-stellar wind boundary, the disk generally remains intact with a relatively stable vertical structure, even when the tilted stellar dipolar corona is rotating relative to the disk.

\subsection{Synthetic Observables}\label{LOS}

AWSoM has the capability to calculate the synthetic Extreme Ultraviolet (EUV) and X-ray emission from the stellar corona in the form of EUV/X-ray images for a given line-of-sight (LOS). These synthetic emissions are designed to match the EUV/X-ray bands that are observed by solar monitors \citep[e.g.,][]{2014ApJ...782...81V}. The flux integral over the images corresponds to the total stellar EUV/X-ray flux at a given time, and the time series of these integrated fluxes provide a light curve of the EUV/X-ray emission, capturing the time variations in these bands due to the disk-corona interaction. The fact that for each snapshot of the simulation we can directly relate the state of the physical solution, and the associated synthetic emission, enables us to relate particular trends in the EUV/X-ray to physical features, such as accretion, heating, or compression episodes.

The top plot in Figure~\ref{fig:3spheres_LOS} shows the time series of the synthetic emission of three EUV bands --- 171, 193, and 304\AA~ \citep[SOHO EIT,][]{1997AdSpR..20.2231D}, together with the AlPoly X-ray band \citep[Suzaku XRT,][]{2007PASJ...59S..23K,2007SoPh..243...63G}. The fluxes of each band are normalized to the initial fluxes at $t=0$h. The time series shows an initial increase in all bands which peaks after 17 hours, which we consider part of the relaxation phase of the simulation. There are two smaller peaks at $t=50$h and $t=78$h, which are clearly seen in the EUV bands (most visible in the 171\AA~band) but are less discernible in the X-ray band.

The bottom three rows of Figure~\ref{fig:3spheres_LOS} show the physical solution during these times, as well as the associated images of the EUV 171\AA~ and the X-ray emission. The left column shows an iso-surface of density for $n=10^9~cm^{-3}$, colored with the sign of the radial velocity --- positive is grey (outflow), and negative is black (inflow). The second column shows an iso-surface of the radial velocity for a value of negative $30~km~s^{-1}$ (inflow), colored with the temperature. At t=17h (the first peak), inflowing material gets heated to over a million K, resulting in a brightening in the EUV/X-ray images, and an overall increase in their fluxes. However, this event is driven by the initial winding of the inner part of the disk during relaxation, and not by an accretion event. The two later peaks in the fluxes might be attributed to moderate episodic heating of infalling material. However, this is coronal material which is moving inwards and is being heated due to a local compression as a result of the disk-corona interaction.

\subsection{Mass and Angular Momentum Transfer}\label{MdotJdot}

In order to estimate mass and angular momentum (AM) transfer in our simulation, we extract a number of spherical shells from the region between the disk and the stellar surface. Unlike some previous simulations \citep[e.g.,][]{2022ApJ...941...73T}, we avoid estimating these parameters exactly at the inner boundary itself, since their values may be affected by the rather complex boundary conditions the model uses. To get a good coverage of the different loss rates at different locations, we estimate these loss rate on sphere at r=2, 3, 5, and 7$R_\star$.

\subsubsection{Mass Transfer}\label{Mdot}

The mass loss rate through a sphere with a radius $r$ is given by the following integral over the sphere:
\begin{equation}
\dot{M}=\int \rho\mathbf{v}\cdot\mathbf{dA},
\end{equation}
where $\mathbf{v}\cdot\mathbf{dA}$ is the component normal to the surface, and the integral is over a sphere of a particular radius (the normal component in this case is simply $v_r$). 

\begin{figure*}[h!]
\centering
\includegraphics[width=5.in]{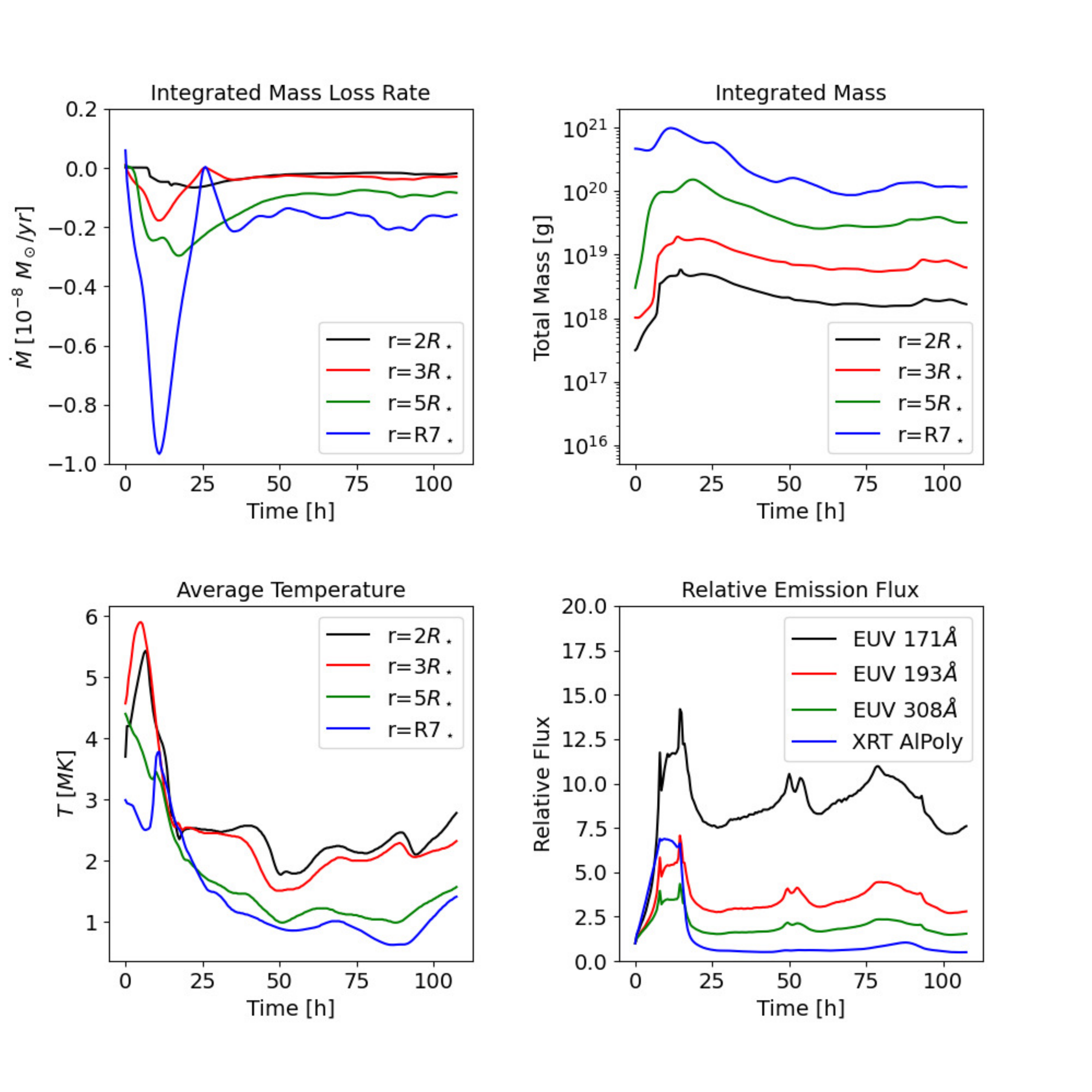}
\caption{Integrated values on four spheres at r=2 (black), 3 (red), 5 (green), and 7 (blue) $R_\star$ as a function of time. The plots are for the mass-loss rate (top-left), integrated mass (top-right), temperature (bottom-left), and integrated synthetic emission flux (bottom-right).}
\label{fig:Lineplots}
\end{figure*}

The top-left panel of Figure~\ref{fig:Lineplots} shows the integrated mass-loss rate over the different spheres as a function of simulated time. Negative values indicates an overall accretion through the sphere while a positive value indicates an overall outward escape of mass. It shows an initially strong event of accretion during the first 20 hours of the simulation through all spheres (weakest at r=2$R_\star$). This initial accretion is associated with the relaxation phase and initial winding of the disk, leading to a large amount of mass moving inwards with rates of $10^{-9}-10^{-8}~M_\odot~yr^{-1}$. 

After about 30 hours, the mass-loss rates at r=5$R_\star$ and 7$R_\star$ are set to about $1-2\cdot10^{-9}~M_\odot~yr^{-1}$, indicating some constant accretion. The accretion rate is almost zero closer to the star, at r=2-3$R_\star$. A similar trend is shown in the top-right panel, which shows the integrated mass over the spheres. The integration is done over the spheres multiplied by a small radial distance of $0.05R_\star$ (the smallest grid size) to provide a volume integral over the mass density, which results in a total mass with units of grams. The total mass plot shows an overall initial accumulation of mass over all spheres, which then declines a bit to a more or less steady value, which is higher than the initial mass on each sphere. This can be explained by the combination of significant mass that is added to the steady state when the disk is superimposed, and the compression of the corona as a result, leading to higher coronal densities. 

The bottom of Figure~\ref{fig:Lineplots} shows the integrated temperature over the spheres (bottom-left), and the associated EUV/X-ray emissions (bottom-right). In both plots, there is an initial increase in the first 20 hours. This increase is clearly associated with relaxation and the compression of the corona by the initial winding of the disk. The increase in density compresses the magnetic field and its magnitude. As a result, the Alfv\'enic energy (which depends on $B^2$) increases and heats the plasma, a heating that leads to an increase in the EUV/X-ray emissions. 

\subsubsection{Angular Momentum Transfer}\label{Jdot}

The angular momentum (AM) loss rate (i.e. torque) is given by the integral over the sphere of the flux of dynamic and magnetic AM. The two components are:

\begin{equation}
\dot{J}_{dyn}=R\int_R  (\rho v_r v_\phi) dA,
\end{equation}
and
\begin{equation}
\dot{J}_{mag}=-R\int_R  \frac{B_r B_\phi}{4\pi} dA,
\end{equation}
respectively, and the total torque is given by
\begin{equation}
\tau=\dot{J}_{dyn}+\dot{J}_{mag}.
\end{equation}

\begin{figure*}[h!]
\centering
\includegraphics[width=5.in]{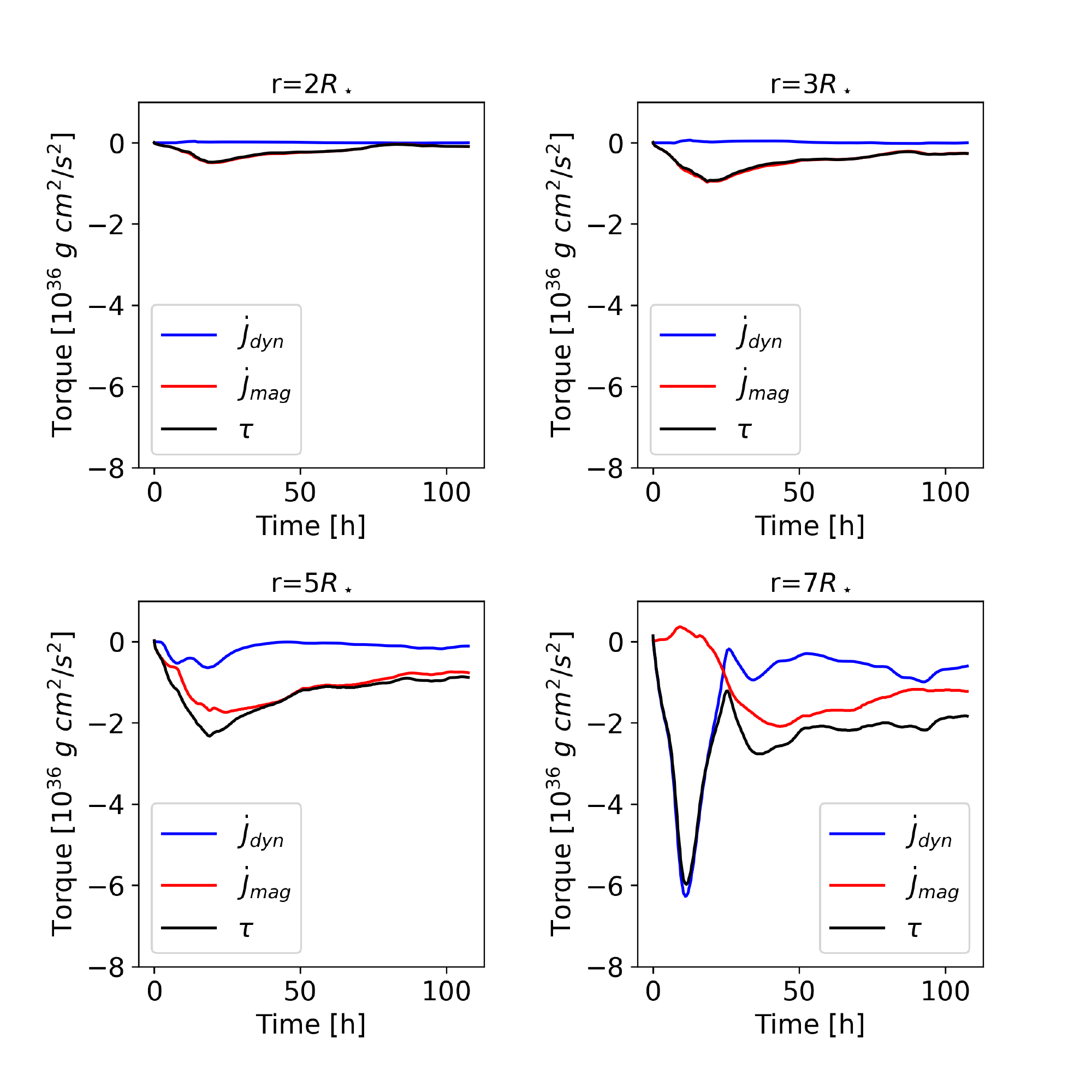}
\caption{The different components of the torque - dynamic (blue), magnetic (red), and total (black) as a function of time. The tores are integrated over spheres at r=2$R_\star$ (top-left), r=3$R_\star$ (top-right), r=5$R_\star$ (bottom-left), and r=7$R_\star$ (bottom-right).}
\label{fig:Jdotplots}
\end{figure*}

Figure~\ref{fig:Jdotplots} shows the time evolution of the different torque components on the different spheres. Positive and negative values indicate an outward or inward transport of AM, respectively. It can be seen that close to the star, at r=2 and 3$R_\star$, the total torque is dominated by the magnetic stresses, which lead to an inward transfer of AM. Further out, at r=5$R_\star$ and especially at 7$R_\star$, the initial winding of the disk lead to a strong inward transfer of AM, which is dominated by the mass flux and dynamic torque. However, after about 30h, the time evolution of the AM at all radii is set to to a more moderate inward AM transfer, dominated by the magnetic torque. Temporal variations are more visible further from the star (at r=7$R_\star$) due to the proximity to the disk-corona interaction region. Overall, inward AM transfer is of the order of $10^{36}~g~cm^2~s^{-2}$, orders of magnitude larger than the AML taken outward by the stellar wind, which is of the order of $10^{29}-10^{32}~g~cm^2~s^{-2}$ \citep[e.g.,][]{2012ApJ...754L..26M,2014ApJ...783...55C}

%%%%%% Discussion %%%%%%%%%%

\section{Discussion} \label{sec:discussion}

\subsection{Accretion Pattern Between the Disk and the Star}\label{AccretionPattern}

\begin{figure*}[h!]
\centering
\includegraphics[width=4.in]{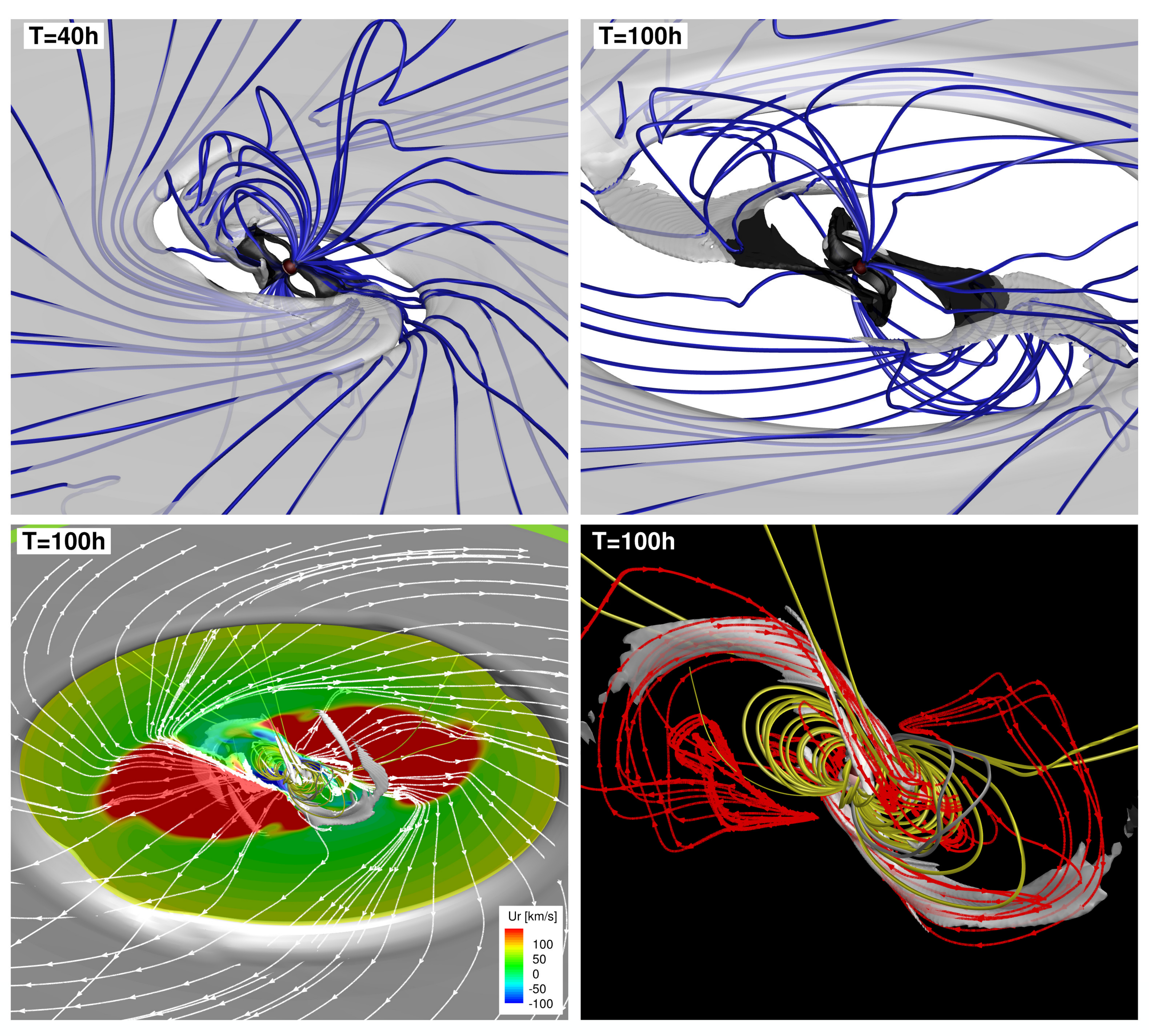}
\caption{Top: density iso-surface of $n=10^{10}~cm^{-3}$ along with selected magnetic field lines shows the region close to the star at t=40h (left), and t=100h (right). Bottom-left: iso-surface of $n=5\times 10^{10}~cm^{-3}$ along with a $z=0$ slice colored with the radial velocity contours at t=100h. White lines represent selected velocity streamlines. Bottom-right: a zoom at the vicinity of the star at t=100h shows selected coronal magnetic field lines in yellow, selected velocity streamlines in red, and an iso-surface of $v_r=-30~km~s^{-1}$. }
\label{fig:streamstructure}
\end{figure*}

The general evolution of our simulation shows an initial winding of the disk close to the star after the disk is superimposed. This is the relaxation part of the simulation, but is interesting because to some extent it resembles an accretion episode in terms of the mass accretion, angular momentum transfer, coronal heating, and an increase in the line emission. This initial relaxation stage lasts for about 25 hours.  As this initial stage completes, the coronal parts which have been eliminated by the insertion of the disk begin to recover, pushing the disk away and reducing significantly the disk mass accretion. The initial cold disk material is slowly pushed outwards by the corona to a greater distance from the star, until the disk-corona boundary is set beyond $10R_\star$. This evolution is shown in Figure~\ref{fig:streamstructure} (as well as in Figure~\ref{fig:global_3d}). 

The main takeaway from these figures is that there is no indication for any infalling material from the disk (black regions do not extend to the disk), and the velocity streamlines that connect the disk and the inner parts of the simulation only go outward. The left and middle panel of Figure~\ref{fig:global_3d} show that the inward radial motion (i.e., accretion), marked by the black region of the density iso-surface is not extended all the way to the disk, but it is associated with coronal material that moves inward. Moreover, there is no strong evidence for a continuous accretion from the disk along the coronal loops or any field line.   

Figure~\ref{fig:Lineplots} shows the time evolution of the mass transfer, temperature, and line emissions. Interestingly, the mass and mass-loss plots do not provide much insight about the source of the peaks at t=50, 78h in the emissions plot. Moreover, the peak at t=50h is associated with a decline in the temperature close to the star (r=2,3$R_\star$), while the peak at t=78h is associated with an increase in the temperature close to the star. It is likely that these variations in the coronal synthetic emissions are due to the change of the overall coronal structure as viewed from the line-of-sight (LOS) of these images as the corona and its tilted dipole field are rotating (and the LOS remains fixed). 

\begin{figure*}[h!]
\centering
\includegraphics[width=5.in]{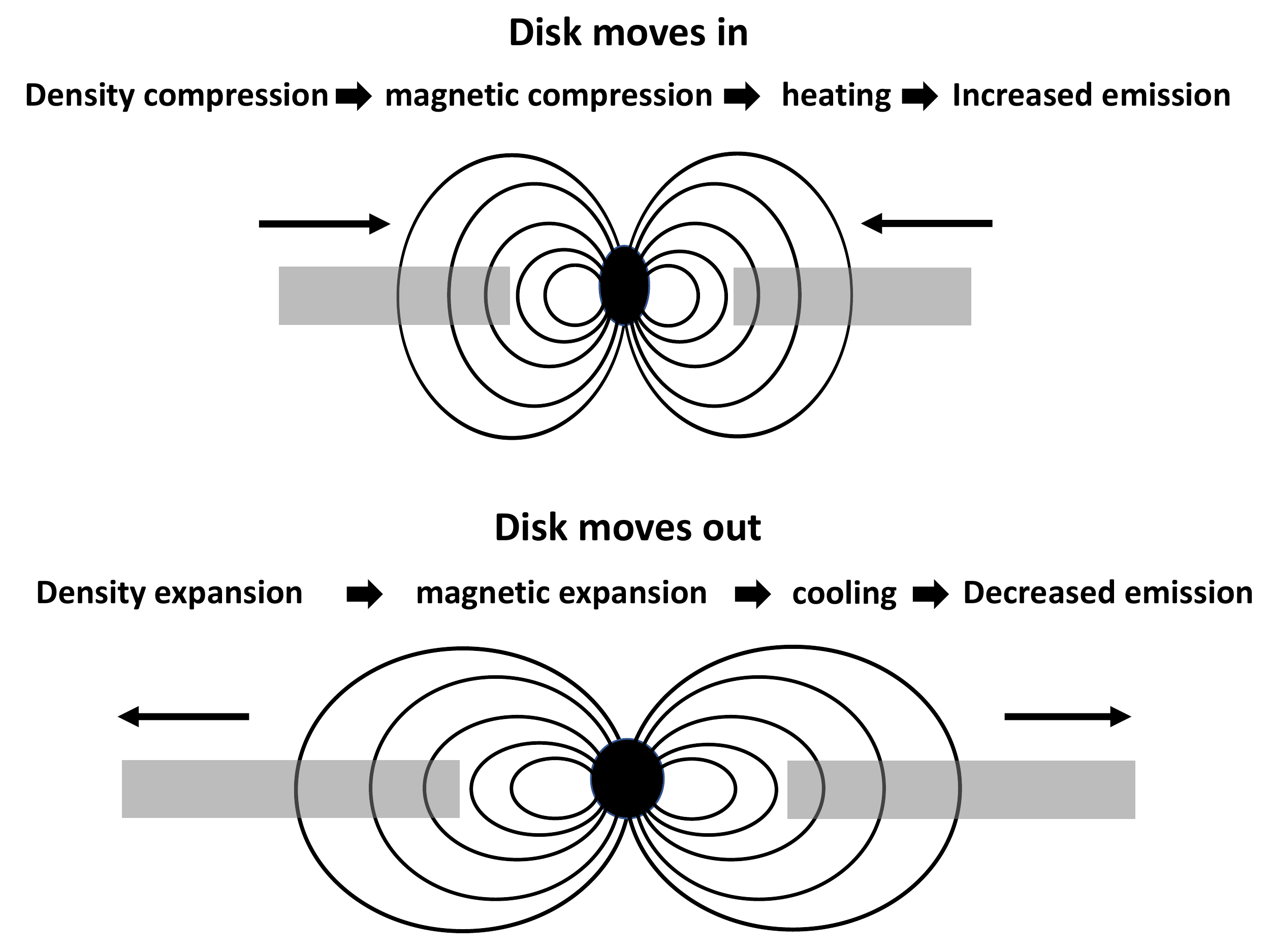}
\caption{Compression (top) and extraction (bottom) of the corona, and the impact on line emission is demonstrated by two cartoons.}
\label{fig:CoronalVariations}
\end{figure*}

Our simulation does not show a clear connection between mass accretion events, and temporal coronal heating and increase in line emission, such as those observed on the Sun in the EUV/UV bands during a failed CME event \citep{2014ApJ...797L...5R,2016A&A...592A..17I}. Nevertheless, our simulation clearly shows a temporal variation in the coronal emission due to coronal compression as a result of the disk-corona interaction. Thus, it is possible that observed temporal variations of the inner parts of protoplanetary disks are driven by pressure/density variations of the corona, and not by episodic mass accretion. This process is shown in Figure~\ref{fig:CoronalVariations}. When the disk pushes inwards, the corona contracts and its density (and mass) increases, leading to an increase in the coronal line emission. When the disk is pushed outwards, the coronal density and mass decrease, leading to a decrease in its line emission. Specifically in our model, plasma compression leads to compression of the magnetic field and an increase in the field strength. As a result, coronal heating increases from the increase in magnetic energy, leading to an increase in the line emission. This effect may not be visible in a model that lacks the magnetic-dependent coronal heating terms.

Overall, after sufficient time, the corona pushes the disk away, without letting a significant amount of mass to accrete. The time variations of the interaction are controlled by the pressure balance variations, and episodic magnetic disconnection events. However, in our simulation, even the latter does not show a strong accretion associated with it in our simulation.

\subsection{Comparing our Results with Previous Models and Observations}\label{AccretionPattern}

A wealth of 2D simulations \cite[e.g.,][]{2004ApJ...615..921M,2008ApJ...678.1109M}, and 2.5D simulations \citep[e.g.,][]{1997ApJ...489..199G,2009A&A...508.1117Z,2017ApJ...845...75B,2017A&A...600A..75B,2020A&A...643A.129P,2022ApJ...929...65I} \citep{2019A&A...624A..50C} investigated the transfer of mass and angular momentum from the disk to the star, and the different accretion patterns. A significant fraction of the three-dimensional MHD studies of disk-magnetosphere interaction in T Tauri systems was done using a model that was initially presented in \cite{2002ApJ...576L..53K} and \cite{2002ApJ...578..420R}. The model has been later used in a series of (2.5, and 3D) follow-up simulations \citep[e.g.,][]{2004ApJ...610..920R,2005ApJ...634.1214L,2008MNRAS.386..673K,2012NewA...17..232L,2019MNRAS.485.2666R,2021MNRAS.506..372R}. This long-term investigation obtained different features of the disk-magnetosphere interaction, including accretion funnels (``tongues") between the disk and the star, hot spots on the surface, Rayleigh–Taylor (RH) and KH instability patterns, and associated prediction for the observed variability of the system. Recently, a three-dimensional model by \cite{2022ApJ...941...73T} demonstrated that the overall torque does not depend on the stellar spin, in contrast to previous studies. To conclude that, the authors distinguished the different components of the single-fluid modeled plasma (e.g., the disk gas, the stellar wind, the disk wind) by calculating their specific entropies, and used this definition to separate the relative contribution of the mass loss/accretion and torque. We argue that this methodology to separate between the different plasma components in the model is uncertain, and that a multi-fluid approach is needed to follow each of the different fluids in a clear manner. In any case, we avoid adopting this approach when analyzing our results. 

All previous models for disk-magnetosphere interaction did not include a self-consistent coronal heating and stellar wind acceleration. At most, a Parker-like stellar wind initial condition is used in the domain, and it is maintained by maintaining high pressure at the inner boundary. It is now known from solar wind observations that a pure Parker wind does not represent the actual solar wind. Not only the fast solar wind requires an additional source of acceleration to explain the observed speed, there is an observed inverse relation between the freeze in solar wind electron temperature and the solar wind speed, indicating an inverse relation between the temperature (and pressure) at the base of the solar wind streamline and the final solar wind speed (inverse to Parker's model) \citep{2007RvGeo..45.1004M}. Recent Parker Solar Probe have shown that this trend is smeared out beyond 0.3~AU \citep{2020ApJS..246...62M}. Therefore, the approach of imposing a Parker solar wind solution on top of a potential stellar field cannot obtain a realistic solar, nor stellar wind. For example, \cite{2004ApJ...615..921M} show velocities of less than $600~km~s^{-1}$, consistent with solar wind simulations that have taken a similar approach, which seems to lack the ability to obtain fast solar wind of $800-900~km~s^{-1}$. 

The setup of the model by \cite{2002ApJ...576L..53K} (as in all the simulations that used it) is even more simplified. It assumes an initially balanced corona-disk system, which define the initial boundary between them. The reference parameters are defined by the stellar mass, radius, and magnetic field via the relations between the Keplerian and the Alfv\'en velocities. Initially, the pressure, density, and temperature have constant but different values in the corona and the disk, respectively, and then the simulation begin to evolve, assuming a relative motion between the disk and the corona. It is important to note that this model is nearly ideal MHD, where it only accounts for the disk viscosity. It has no coronal heating and wind acceleration terms, as well as thermodynamics terms, which our model includes. Thus, while capturing some MHD effects, the results of this model cannot capturing any of the processes that are missing, which seem to be important for the disk-corona interaction.

Our simulation shows that the inclusion of the self-consistent stellar corona and stellar wind reduce the disk accretion as well as the transfer of angular momentum between the disk and the star \citep[as also shown by, e.g.,][]{2017ApJ...845...75B}. Overall, the inward transport of angular momentum is dominated by the magnetic torque, not the dynamic one. After the initial superposition of the disk on top of our coronal solution, the inclusion of the self-consistent terms in our model lead to a rebuild of the corona and the pushing of the disk outwards. Overall, these terms lead to a pressure balance between the disk and the corona, which is different from the one obtained by the models that do not include them. Finally, since our model includes the stellar wind solution, the interaction between the wind and the disk is well captured, enabling to analyze the magnitude of the vertical disk evaporation and the stability of the boundary between the disk and the stellar wind \citep[this was also investigated recently by][]{2021MNRAS.507.1106C}. 

\subsection{Non-vertical magnetic field in the disk and the Growth of Azimuthal Field}\label{NonVerticalB}

\begin{figure*}[h!]
\centering
\includegraphics[width=7.in]{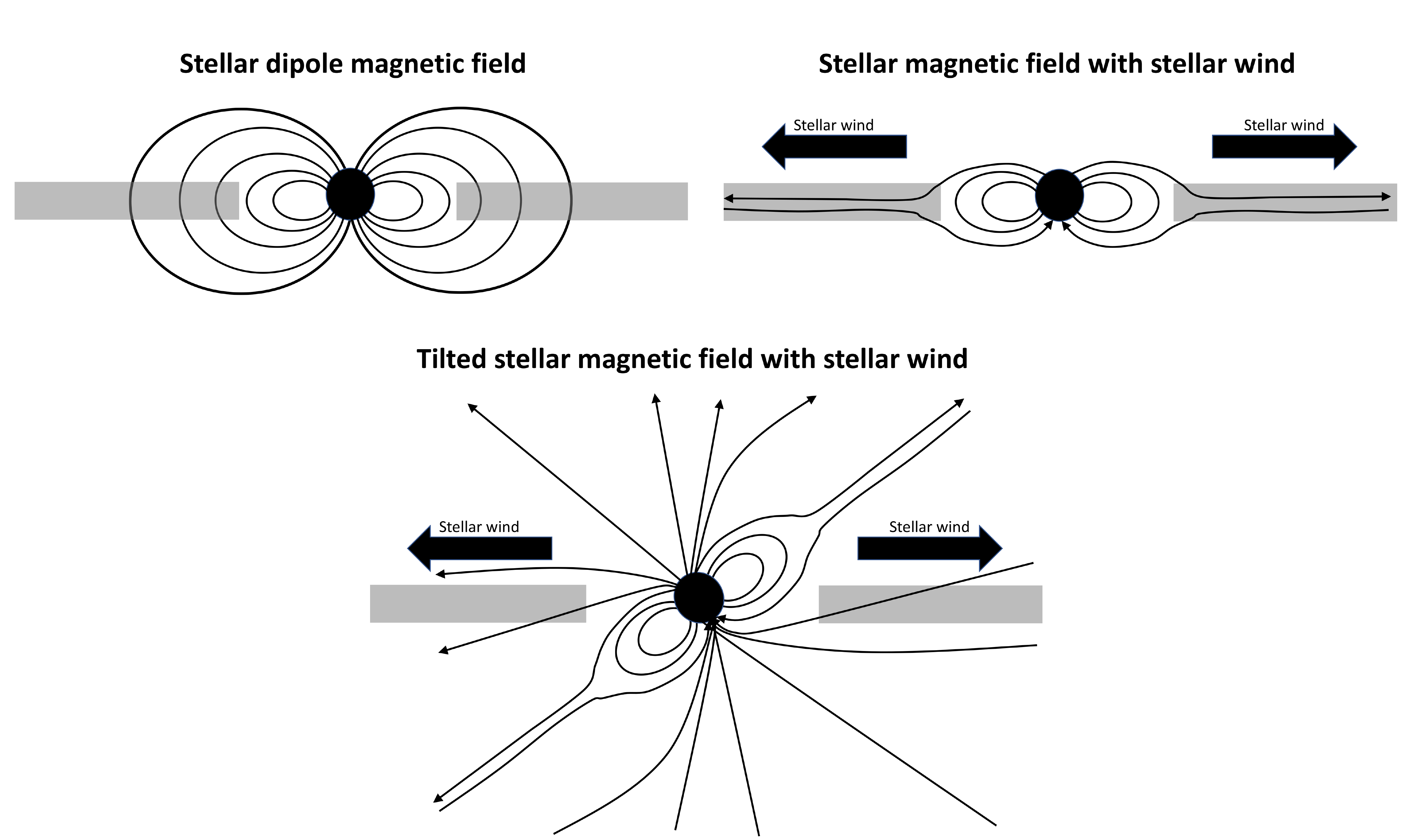}
\caption{Top-left: a pure stellar dipole field extends into the disk with a vertical field near the equator. Top-right: the existance of a stellar wind stretches the magnetic field and open it up at the Alfv\'en point. Thus, the field near/inside the disk is expected to be radial, not vertical. If it si affected by the disk rotation, it is azimuthal, but still not vertical. Bottom: if the stellar dipole field is tilted, the field is still mostly radial.}
\label{fig:DipolarStellarField}
\end{figure*}

\begin{figure*}[h!]
\centering
\includegraphics[width=5.in]{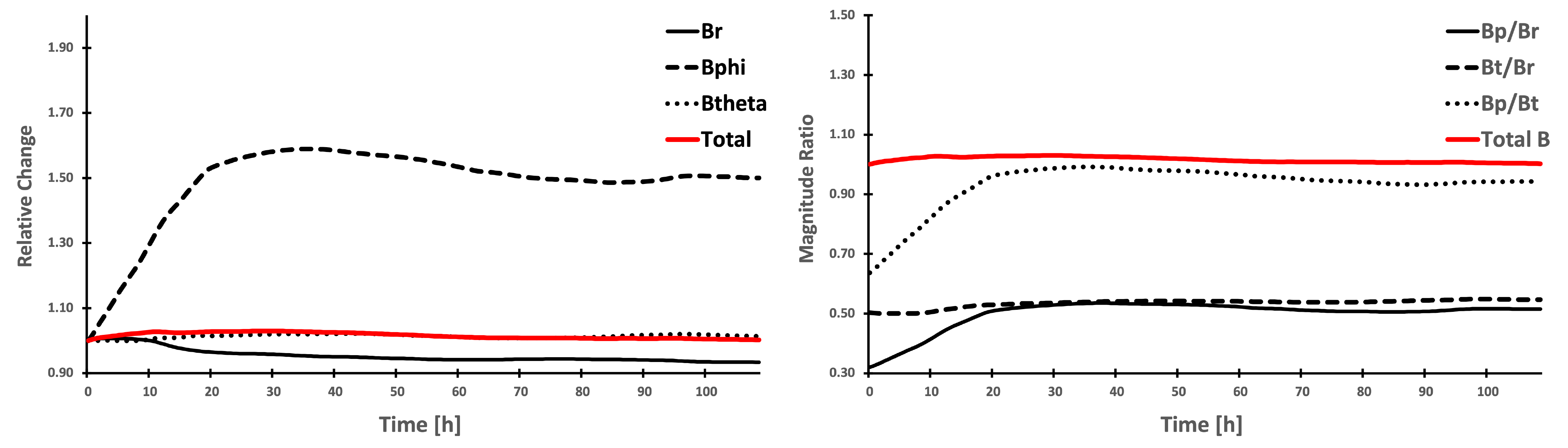}
\caption{Time evolution of the integrated different components of the magnetic field on the equatorial plane. Left: the individual components of the field - radial ($B_r$, solid black), vertical ($B_\theta$, dashed black), azimuthal ($B_\phi$, dotted black), and total (solid red). Right: the ratio of the individual components. }
\label{fig:Bchange}
\end{figure*}

One important feature that our simulation shows is a clear non-vertical magnetic field in the disk. Traditional views assume a dipole-like stellar field (top-left panel of Figure~\ref{fig:DipolarStellarField}), which extends towards large distances, so that the assumed ``background" magnetic field in a small part of the disk is vertical. This initially vertical field is assumed in many calculations/simulations of magnetic dynamo and MRI growth in a small box of the disk \citep[e.g.,][]{1991ApJ...376..214B,2020apfs.book.....A}, as well as in global disk-magnetosphere simulations \citep[e.g.,][]{2017ApJ...845...75B,2017A&A...600A..75B,2020A&A...643A.129P}. In reality, the stellar wind stretches and opens up the stellar field so the field is nearly radial in the interplanetary space, including the vicinity of the disk (top-right panel of Figure~\ref{fig:DipolarStellarField}). The same will apply to a tilted-dipole stellar field (bottom panel of Figure~\ref{fig:DipolarStellarField}). If the disk winds up the field due to its Keplerian motion (assuming the disk is sufficiently ionized), then the disk will have an azimuthal component as well. Our simulation shows this growth in azimuthal field (see Figure~\ref{fig:Bchange}). However, this view predicts a radial/azimuthal background field around the disk, where a strong vertical field is less likely. While this field geometry still allows for the growth of MRI, it may operate differently, so the alternative background field geometry should be considered. 

%%%%%% Conclusions %%%%%%%%%%

\section{Summary \& Conclusions} \label{sec:summary}

We present a three-dimensional simulation of the short-term interaction of protoplanetary disk and the stellar corona. Our simulation includes the disk, the stellar magnetic field, and self-consistent coronal heating and stellar wind acceleration. Our main findings are:
\begin{enumerate}
\item At the beginning of the simulation when the system relaxes from its artificial initial conditions (the first 25h), the inner part of the disk winds around and moves very close to the star in what seems to be a strong accretion. However, after about 30h, the corona begin to build up its original state. This pushes the disk out until the disk-corona boundary is beyond $10R_\star$. 
\item A strong, steady disk accretion which is funneled along coronal field lines is not clearly visible at any stage of the simulation. Instead, very weak, sporadic accretion is observed.
\item We produce synthetic line emission lightcurves which show flare-like increases during the time of the simulation. These flux increase events are not correlated with accretion events nor with heating events. Our simulation points to variations in the line emission flux being the result of coronal compression and extraction due to the disk-corona pressure variations.
\item Disk-wind evolves above and below the disk. However, the disk - stellar wind boundary stays quite stable, and any disk material that reaches the stellar wind region is advected out by the stellar wind.
\end{enumerate}

The simulation presented here is the first attempt to simulate protoplanetary systems with AWSoM. Future simulations may include the extention of the simulation domain to greater distances, using an arbitrary stellar magnetic field, the inclusion of a neutral fluid, and the inclusion of ionization processes in the disk. 

\begin{acknowledgments}
We thank an unknown referee for their useful comments that improved this manuscript. This work is supported by NASA XRP grant 80NSSC20K0840 and {\it Chandra} grants TM6-17001A and TM9-20001X. Simulation results were obtained using the (open source) Space Weather Modeling Framework, developed by the Center for Space Environment Modeling, at the University of Michigan with funding support from NASA ESS, NASA ESTO-CT, NSF KDI, and DoD MURI. The simulations were performed on NASA's Pleiades cluster (under SMD-20-52848317) and on the Massachusetts Green High Performance Computing Center (MGHPCC) cluster.
\end{acknowledgments}

%%%%%% Bibliography %%%%%%%%%%%%%%%%

%\bibliography{DiskBib}{}
%\bibliographystyle{aasjournal}

%%%%%% Figures %%%%%%%%%%%%%%%%

\begin{figure*}[h!]
\centering
\includegraphics[width=6.in]{LOS_fluxes.pdf}
\caption{Top panel: integrated fluxes of the synthetic EUV/X-ray images produced by the simulation as a function of time. The fluxes are normalized to their individual initial fluxes at t=0 hours. Dashed lines mark the temporal peak increase in the fluxes at t=17, 50, and 78 hours. Bottom three panels show the synthetic 171\AA~EUV and XRT images (two left columns, respectively) during the time of the peak flux increase. The left panel shows an iso-surface similar to that in Figure~\ref{fig:global_3d}. The second column shows an iso-surface of $v_r=-30~km~s^{-1}$, which represents an infalling material region, colored with contours of the temperature. It shows some, but very little cold material accreting. The left two columns show the three-dimensional structure close to the star during the the time of the EUV/X-ray flux increase and from the same viewing angle as the LOS of the images in the two right columns.}
\label{fig:3spheres_LOS}
\end{figure*}

\end{document}